\documentclass{jpsj-suppl}
\usepackage{txfonts}

\title{
{\it Ab initio} Disordered Local Moment Approach
for a Doped Rare-Earth Magnet}

\author{Munehisa~\textsc{Matsumoto}$^{1}$,
Rudra~\textsc{Banerjee}$^{2}$,
and Julie~B.~\textsc{Staunton}$^{2}$}

\inst{$^{1}$
Elements Strategy Initiative Center
for Magnetic Materials (ESICMM),\\
National Institute for Materials Science,
Tsukuba 305-0047, JAPAN\\
$^{2}$
Department of Physics, University of Warwick,
Coventry CV4 7AL, United Kingdom}

\email{MATSUMOTO.Munehisa@nims.go.jp}

\recdate{October 18, 2014}

\abst{Following the finite-temperature {\it ab initio} calculation framework
based on the relativistic disordered local moments~\cite{julie},
we computationally demonstrate
the possibility of doping-enhanced coercivity at high-temperatures,
taking YCo$_5$ as a working material
in order to extract the $3d$-electron part
of the electronic structure of the rare-earth permanent magnets.
Alkaline-earth dopants are shown to be the
candidates to realize the proposed phenomenon.}

\kword{
rare-earth magnets, first principles, relativistic disordered local moment, magnetic anisotropy}

\begin{document}
\maketitle

\section{Introduction}

\subsection{Motivation}

\begin{figure}
\begin{center}
\scalebox{0.6}{
\includegraphics{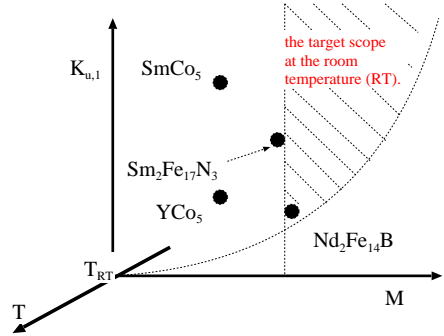}
}
\end{center}
\caption{\label{fig::target_scope} Schematic picture to show
the target scope for the next-generation permanent magnet
and the status
of the working material in the present calculation. Even though the magnetic anisotropy can show
a non-monotonic temperature dependence for some rare-earth magnets~\cite{hirosawa_1986}, the target
scope is practically drawn with a set of representative numbers taken at the room temperature.}
\end{figure}

Demands for an economically robust way to improve
the merits of rare-earth permanent magnet materials
at high temperatures has been recently sought~\cite{msj_2012}.
One of the most important merits is the coercivity $H_{\rm c}$, which
is the resistance of magnetization
against the externally applied reverse-direction magnetic field. Coercivity seems to
be realized by both of the extrinsic factors such as the microstructure and the intrinsic factors
such as the strength of the magnetic anisotropy~\cite{msj_2012}. In the present work
we focus on the latter which can be addressed within solid state physics, and give
{\it ab initio} results for magnetization $M$ and the uni-axial
magnetic anisotropy energy (MAE) $K_{\rm u1}$ at finite temperatures. Our data
can meet the demand for coercivity under the empirical relation
that $H_{\rm c}$ is roughly proportional to the anisotropy field,
\[
H_{\rm A}=\frac{2K_{\rm u1}}{M}.
\]
The merit of the intrinsic properties of
permanent magnets is characterized on a two-dimensional parameter
space spanned by $K_{\rm u1}$ and $M$
as schematically shown in Fig.~\ref{fig::target_scope}. 

To put more specifically the problem, we wish to find a way to
enhance the high-temperature coercivity of today's champion magnet Nd$_2$Fe$_{14}$B without resorting to
Dy doping, which is thought to work mostly to enhance the ground-state anisotropy field~\cite{hirosawa_1986}.
Among the magnetically-relevant electrons in rare-earth permanent magnets which are
\begin{itemize}
\item $4f$-electrons localized on the rare-earth atoms,
\item $3d$-bands from the transition metals, and
\item $5d$-bands from the rare-earth that couples the above two, 
\end{itemize}
the high-temperature tail of the magnetic anisotropy is
mostly carried by $3d$-electrons~\cite{skomski_1998}.
We propose a way to enhance the high-temperature anisotropy field
by engineering the $3d$-electrons,
which could potentially lead to an alternative scheme to Dy-doping. 
We take the case of YCo$_5$ which represents one of the simplest $3d$-electron physics
in rare-earth permanent magnets and implement the $3d$-band
engineering by a hole doping~\cite{mm_2014}.

\subsection{The target material YCo$_5$}

\begin{figure}[tbh]
\begin{center}
\begin{tabular}{ll}
(a) & (b) \\
\scalebox{0.7}{
\includegraphics{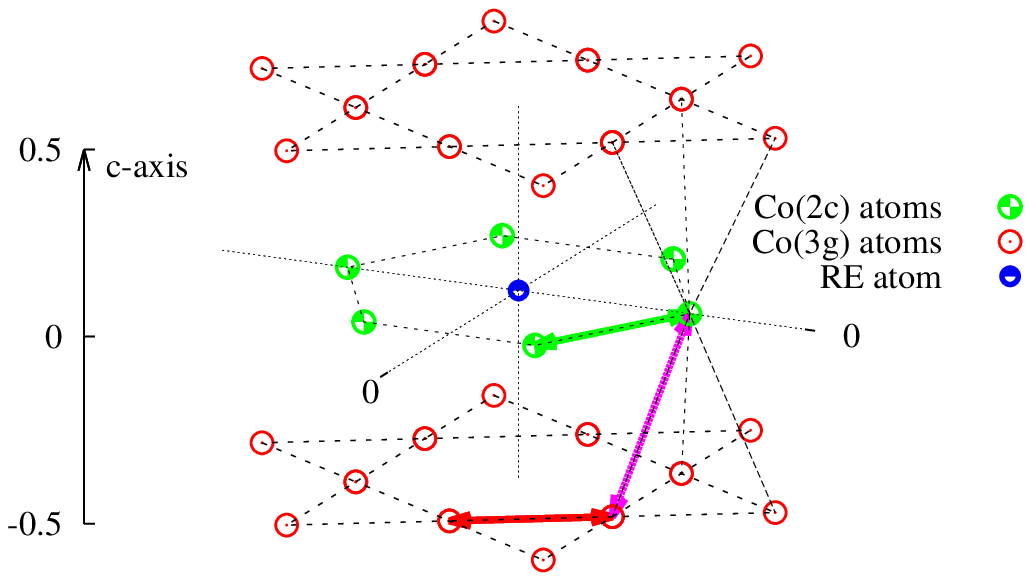}}
&
\scalebox{0.55}{
\includegraphics{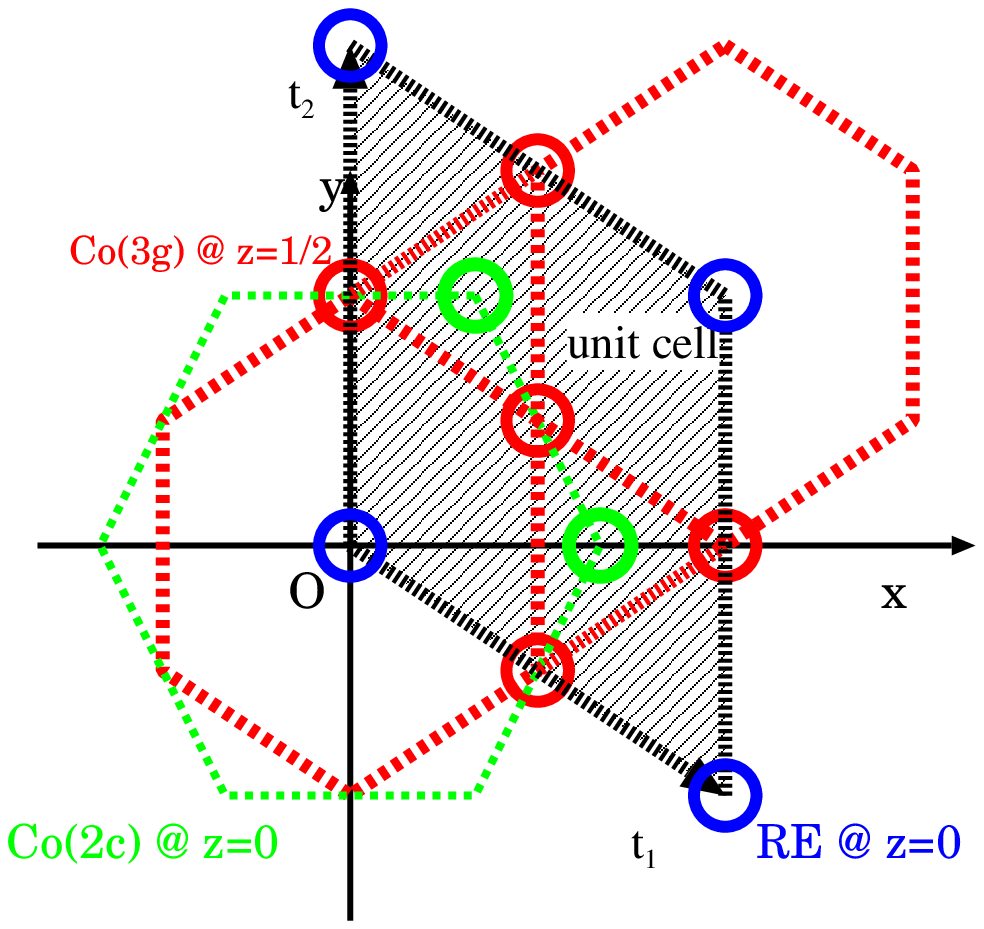}}
\end{tabular}
\end{center}
\caption{Crystal structure of YCo$_5$ (a) with
a bird's eye view and (b) as projected onto the $ab$-plane. We have set the $z$-axis
along the crystallographic $c$-axis.}
\label{fig::crystal_structure}
\end{figure}
The permanent magnet material YCo$_5$ consists of the Co(2c) honeycomb lattice
along the $ab$-plane, of which center the rare-earth (RE) (here Y) resides,
Co(3g) kagom\'{e} lattice along the $ab$-plane, and
Co(2c)-Co(3g) almost regular tetrahedra that are chained along the $c$-axis
as shown in 
Fig.~\ref{fig::crystal_structure}~(a)
with the plot for the unit cell
as projected onto the $ab$-plane in Fig.~\ref{fig::crystal_structure}~(b).

\subsection{Methods}

We follow {\it ab initio} relativistic disordered local moment approach (DLM)~\cite{julie}
based on Korringa-Kohn-Rostoker (KKR) method and coherent-potential approximation (CPA)
to address the MAE at finite temperatures. The basic idea
is to express the free energy of a given material as
\[
F^{(\hat{\mathbf n})}=F_{\rm iso}+K_{\rm u 1}\sin^{2}\theta
\]
where $F_{\rm iso}$ is the isotropic part,
$K_{\rm u 1}>0$ is the MAE for uni-axial MCA
and $\theta$ is the angle between the direction of magnetization
and the easy axis. For a uni-axial magnet,
we fix the direction
of magnetization to be $\hat{\mathbf n}=(1,0,1)/\sqrt{2}$
that is, $\hat{\mathbf n}=(\sin\theta\cos\phi,\sin\theta\sin\phi,\cos\theta)$ with $\theta=\pi/4$
and $\phi=0$, and calculate the magnetic torque
\[
T_{\theta}\equiv -\frac{\partial F}{\partial\theta}=-2K_{\rm u1}\sin\theta\cos\theta
\]
with $T_{\theta=\pi/4}$ to get
\[
K_{\rm u 1}=-T_{\theta=\pi/4}.
\]
The full details of the free energy in DLM are described in Ref.~\cite{julie} and the sketch of them
is given in Sec.~II~A in Ref.~\cite{mm_2014}.

\section{Doping-induced enhancement of the finite-temperature
anisotropy field}

\subsection{The strategy}

\begin{figure}[tbh]
\begin{center}
\begin{tabular}{ll}
(a) & (b) \\
\scalebox{0.6}{
\includegraphics{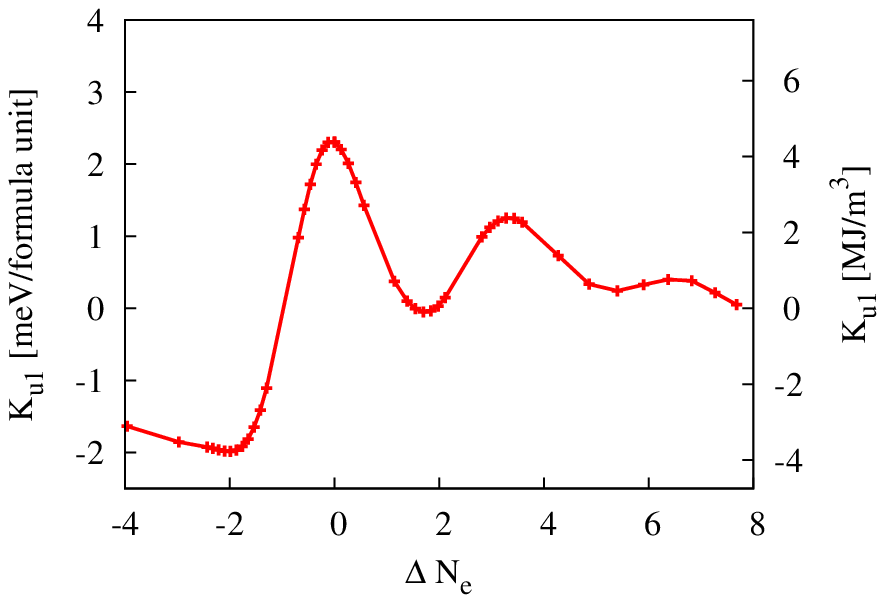}}
&
\scalebox{0.6}{
\includegraphics{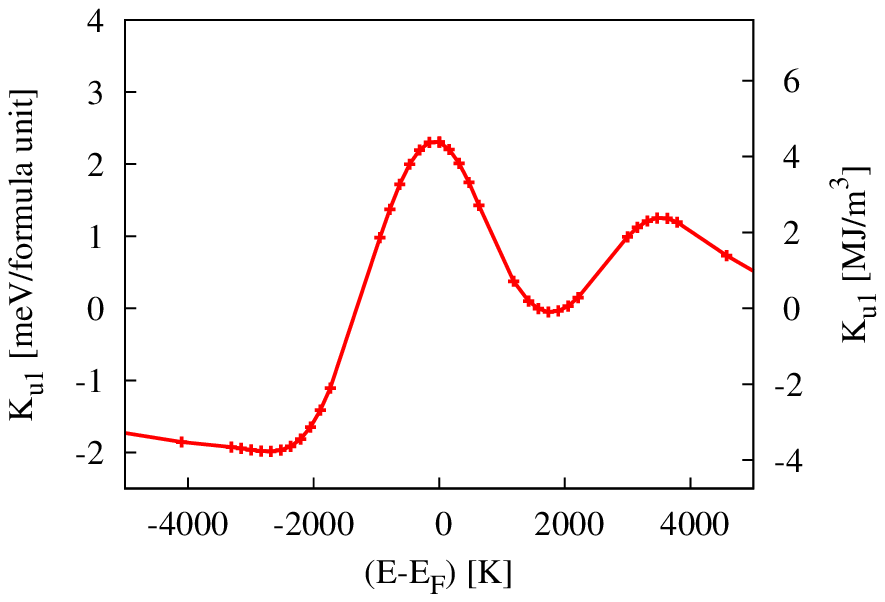}}
\end{tabular}
\end{center}
\caption{(a) Calculated filling dependence of
MAE for YCo$_5$ near $T=0$. The filling is measured with the electron number
at the Fermi level (which is 54, including Y($4p$)-band in the valance states)
set to be zero. (b) The same thing is plotted as a function of the
relative energy as measured in Kelvins from the position of the Fermi level.}
\label{fig::MAE_T_sim_0}
\end{figure}
Calculated magnetic anisotropy energy $K_{\rm u1}$
for the undoped YCo$_5$ near the ground state
is shown in Fig.~\ref{fig::MAE_T_sim_0}~(a) as a function of
the number of valence electrons. The horizontal axis is plotted
relatively to the valence electron number right on the Fermi level,
which is $N_e=54$. For the calculations of hole doping effects,
we include Y($4p$) states in the valence states
to add up 6 electrons to $N_e=48$. The latter number is indeed the standard
one found in the literature~\cite{igor_2003_prb} for the undoped case.
We observe a peak structure which seems to
originate in filling up an electronic state
which carries the magnetic anisotropy. Now the idea is to focus on
the particle-hole excitations that happens near the Fermi level $\Delta N_e=0$
at finite temperatures. That would just lead to the temperature
decay of $K_{\rm u1}$ on $\Delta N_e=0$ while if the filling is
fixed to be $\Delta N_e\stackrel{<}{\sim}0$, finite temperature effects could
enhance $K_{\rm u1}$ by filling up the electronic state that correspond to
the peak in Fig.~\ref{fig::MAE_T_sim_0}~(a).

In order to figure out the distance to the peak position as measured
in terms of the temperature scale, the same data for
$K_{\rm u1}$ is shown in Fig.~\ref{fig::MAE_T_sim_0}~(b) as a function
of the manually-swept Fermi energy as measured from the one for the undoped case.
The scale of the typical operating temperature range for permanent magnets,
$200~\mbox{[K]}\stackrel{<}{\sim}T\stackrel{<}{\sim}500~\mbox{[K]}$, is seen to correspond
to the filling variation of a few of $0.1$ electrons. Thus we set up
computational sample materials to dope 0.1 or 0.2 holes in the unit cell of
YCo$_5$ by some element replacements
and monitor the temperature dependence of their $K_{\rm u1}$.
Below we discuss two cases, Y$_{1-x}$Ca$_{x}$Co$_{5}$
and YCo$_{5-x}$Cu$_{x}$.

\subsection{Replacing rare-earth with alkaline-earth}

\begin{figure}[tbh]
\begin{center}
\scalebox{0.8}{
\includegraphics{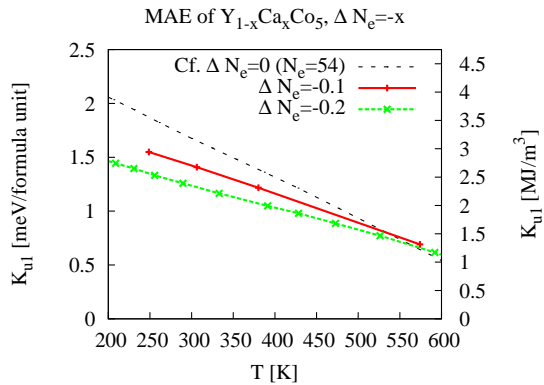}}
\end{center}
\caption{Calculated temperature dependence of MAE for Ca-doped YCo$_5$.}
\label{fig::final_touch}
\end{figure}

One of the standard ways to implement the hole doping has been
to replace tri-valent
rare-earth elements by di-valent alkaline-earth dopants.
Here we replace Y by Ca in YCo$_5$ and calculate
the temperatute dependence of the uni-axial MAE, $K_{\rm u1}$.
The results are shown in Fig.~\ref{fig::final_touch}. Indeed some
temperature enhancement is numerically observed
in the high-temperature
data at $T>550~\mbox{[K]}$ for the doping ratio $x=0.1$ and $0.2$
in Y$_{1-x}$Ca$_{x}$Co$_{5}$.

\paragraph{Remarks on YCo$_{5-x}$Cu$_{x}$}
It is expected that Cu dopes electrons rather than holes when it replaces Co
in YCo$_5$. Indeed
doping Cu shifts the calculated Fermi level upward and computatinal
samples indeed have more electrons, but we have numerically observed
that the peak position of MAE shifts upward
on the $N_e$ axis
even faster,
thus an effective hole doping effects
seem to be at work for the MAE of YCo$_{5-x}$Cu$_{x}$. However Cu dilutes
the magnetically active $3d$-electrons on Co and the bulk MAE gets just weaker
and weaker with respect to Cu doping. Thus we have not been able to numerically
demonstrate the temperature enhancement of MAE
around such diminished peak.

\section{Conclusions and Outlook}

We have demonstrated
the doping-enhanced coercivity at high temperatures~\cite{mm_2014}
with the real-element dopant Ca onto Y. Alloying
with other alkalline-earth dopants, Sr or Ba, to make
Y$_{1-x}$Sr$_{x}$Co$_5$
and Y$_{1-x}$Ba$_{x}$Co$_{5}$,
would also be the candidate materials
to monitor the $3d$-electron dominated
magnetic anisotropy to demonstrate
the enhancement of high-temperature coercivity.
Holes can be doped
either chemically as we have calculated or 
by a slight application of positive electric voltage,
which
could also be useful for improving the high-temperature coercivity of Nd$_{2}$Fe$_{14}$B.

In the present work, we have focused on a framework for a) {\bf $3d$-electron engineering}.
The next target material would be GdCo$_5$
where we can extract the $5d$-electron part
and discuss b) {\bf $5d$-electron engineering}
as well from first principles. c) {\bf Engineering on
$4f$-electron part} is investigated on the basis of LDA+DMFT~\cite{mm_2009}.
All of the above a)-c) is now being pursued to computationally implement the elements strategy
for permanent magnets.

\paragraph{Acknowledgements} Helpful discussions on the electronic structure
of RCo$_5$ [R=Y, Gd, Sm\ldots]
with T.~Miyake, H.~Kino, H.~Akai, Y.~Yamaji,
M.~Ochi, R.~Arita, and R.~Akashi
are gratefully acknowledged. This work is supported by
ESICMM under the outsourcing project of the Ministry of Education,
Culture, Sports, Science and Technology (MEXT), Japan.
Support is acknowledged from the EPSRC~(UK)~grant~EP/J006750/1~(J.B.S.~and
~R.B.).
Numerical computations were performed on ``Minerva''
at University of Warwick, SGI Altix at National Institute for Materials Science,
and Supercomputer System B in the Institute for Solid State Physics, University of Tokyo.

\end{document}